\documentstyle[11pt, aaspp4]{article} 

\newcommand{\Teff} {T_{\rm eff}}
\newcommand{\logg}{\log g}
\newcommand{\vsini}{v \sin i}
\newcommand{\kms}{\, {\rm km} \, {\rm s}^{-1}}
\newcommand{\Weq}{W_\lambda}
\newcommand{\etal}{{\it et al.\ }}
\newcommand{\e}{$\pm \;$}

\newcommand{\unit}[1]{\, {\rm #1}}

\lefthead{Behr et al.}
\righthead{Abundance Anomalies in M13}

\begin{document}

\title{Striking Photospheric Abundance Anomalies \\ in Blue Horizontal-Branch Stars in Globular Cluster M13\altaffilmark{1}}

\author{ Bradford B. Behr\altaffilmark{2},
	Judith G. Cohen\altaffilmark{2},
	James K. McCarthy\altaffilmark{3},
	S. George Djorgovski\altaffilmark{2} }

\altaffiltext{1}{Based in large part on observations obtained at the
	W.M. Keck Observatory, which is operated jointly by the California 
	Institute of Technology and the University of California}
\altaffiltext{2}{Palomar Observatory, Mail Stop 105-24,
	California Institute of Technology, Pasadena, CA, 91125}
\altaffiltext{3}{PixelVision, Inc., 4952 Warner Avenue, Suite 300, Huntington Beach, CA, 92649}

\begin{abstract}

High-resolution optical spectra of thirteen blue horizontal-branch (BHB) stars in the globular cluster M13 show enormous
deviations in element abundances from the expected cluster metallicity. In the hotter stars ($\Teff > 12000 \unit{K}$),
helium is depleted by factors of 10 to 100 below solar, while iron is enhanced to three times the solar abundance, two orders
of magnitude above the canonical metallicity [Fe/H]~$\simeq -1.5$~dex for this globular cluster. Nitrogen, phosphorus, and chromium
exhibit even more pronounced enhancements, and other metals are also mildly overabundant, with the exception of
magnesium, which stays very near the expected cluster metallicity. These photospheric anomalies are most likely due to
diffusion --- gravitational settling of helium, and radiative levitation of the other elements --- in the stable
radiative atmospheres of these hot stars. The effects of these mechanisms may have some impact on the photometric morphology 
of the cluster's horizontal branch and on estimates of its age and distance.

\end{abstract}



\section{Introduction}

A number of unresolved issues in post-main-sequence stellar evolution revolve around the nature of stars on the
horizontal branch (HB), an evolutionary stage characterized by core helium burning and shell hydrogen burning. The HB
stars in globular clusters are particularly appealing targets, as they are readily identified by their position in a
cluster's color-magnitude diagram, and are assumed to be chemically homogeneous and coeval with the other stars in the
cluster. Although intrinsically luminous, most cluster HB stars are also distant, with V magnitudes of 14 or greater, so
that detailed spectroscopic study is challenging. With the recent advent of 8--10~meter-class telescopes and highly efficient
spectrographs, however, the HBs of many of the nearer globular clusters are now accessible at high spectral resolution in reasonable
exposure times. We have therefore undertaken a program to measure chemical abundances and rotation rates of
HB stars in M3, M13, M15, M92, and M68 via high-resolution echelle spectroscopy.

M13 (NGC 6205) is one of the closest and best-studied globulars, with $(m-M) = 14.35$ mag (Peterson 1993) and a metallicity
[Fe/H]~$= -1.51$~dex measured from red giant abundances (Kraft \etal 1992). Its BHB extends from the blue edge of the RR
Lyrae gap to rather high temperatures (a ``long blue tail''), but is interrupted by one or more gaps, including a large
one at $U-V \simeq -0.3$~mag. Some researchers have suggested that this gap separates two different populations of BHB
stars (Ferraro \etal 1997b; Sosin \etal 1997). In order to test this hypothesis, we have observed stars on either side of the gap,
to look for differences in composition and rotation. In this paper, we describe the striking trends in helium and
metal abundance that we observe along the HB of M13. The rotation results will be reported in a subsequent paper (Behr \etal 1999a).


\section{Observations and Reduction}

The spectra were collected using the HIRES spectrograph (Vogt \etal 1994) on the Keck I telescope, during four observing
runs on 1998 June 27, 1998 August 20--21, 1998 August 26--27, and 1999 March 09--11. A 0.86-arcsec slit width
yielded $R = 45000$ ($v = 6.7 \kms$) per 3-pixel resolution element. Spectral coverage ran from $3940-5440$~\AA\ ($m =
90-66$) for the June observations, and $3890-6280$~\AA\ ($m = 91-57$) for the August and March observations, with slight gaps above
5130~\AA\ where the free spectral range of the orders overfilled the detector. We limited frame exposure times to 1200 seconds, to minimize
susceptibility to cosmic ray accumulation, and then coadded three frames per star. $S/N$ ratios were on the order of
$50-90$ per resolution element, permitting us to measure even weak lines in the spectra.

Nine of the thirteen stars in our sample were selected from HST WFPC-2 photometry of the center of M13 from Ferraro \etal
(1997a), as reduced by Zoccali \etal (1999). The program stars were selected to be as isolated as possible; the HST images showed no 
apparent neighbors within
$\simeq 5$ arcseconds. The seeing during the HIRES observations was sufficiently good ($0.8-1.0$ arcsec) to avoid any risk
of spectral contamination. The $U-V$ colors from the HST study provided the $\Teff$ estimates for the subsequent
abundance analysis. The other four M13 HB program stars were taken from the $\vsini$ and [O/H] survey of Peterson,
Rood, and Crocker (1995). They are all located in the cluster outskirts, where crowding is not so problematic. 
Positions, finding charts, photometry, and observational details for the target stars will be provided in a later paper (Behr 1999b).

We used a suite of routines developed by J.K. McCarthy (1988) for the FIGARO data analysis package (Shortridge 1988)
to reduce the HIRES echellograms to 1-dimensional spectra. Frames were bias-subtracted, flat-fielded against exposures of
HIRES' internal quartz incandescent lamps (thereby removing much of the blaze profile from each order), cleaned of
cosmic ray hits, and coadded. A thorium-argon arc lamp provided wavelength calibration. Sky background was negligible, and 1-D
spectra were extracted via simple pixel summation. A 10th-order polynomial fit to line-free continuum regions completed
the normalization of the spectrum to unity.


\section{Analysis}

The resulting spectra show many tens to over two hundred metal absorption lines each. Line broadening from stellar
rotation is evident in many stars, but even in the most extreme cases, the line profiles were close to Gaussian, so
line equivalent widths ($\Weq$) were measured by least-square fitting of Gaussian profiles to the data. Equivalent
widths as small as 10 m\AA\ were measured reliably, and errors in $\Weq$ (estimated from the fit $\chi^2$) were typically
5 m\AA\ or less.

Observed lines were matched to the atomic line lists of Kurucz \& Bell (1995). (Several observed lines in the hotter stars could
not be identified, and a more comprehensive future analysis will attempt to do so.) Those lines that were
identified provided a consistent $v_r$ solution for each of the stars, placing all of them well within the canonical
heliocentric $v_r = -246.6 \kms$ of M13.

We make the simplifying assumption that all the program stars lie on or near the zero-age horizontal-branch (ZAHB) track
computed by Dorman \etal (1993), so that surface gravity would be determined by our choice of temperature. For the
hotter and potentially ``overluminous'' HB stars (discussed below), this assumption may overestimate $\logg$ by
as much as $\simeq 0.7$ dex (Moehler 1998, Figure 4), but this will have only a modest impact on
computed abundances for the species observed. Effective temperature was
derived for the nine HST stars by matching dereddened $U-V$ color indices to computed ATLAS9 colors, with
errors in $\Teff$ based on the photometric errors. For the four non-HST stars, we accepted the published $\Teff$ values
of Peterson \etal (1995), although we note that these are based on photographic $B-V$ photometry, and are thus suspect.
Str\"{o}mgren photometry of M13 will refine the $\Teff$ for these stars in a later reanalysis.
We assign conservative error bars of $\pm 500 \unit{K}$ to the photographic temperatures.

For the chemical abundance analyses, we use the LINFOR/LINFIT line formation analysis package (developed at
Kiel, based on earlier codes by Baschek, Traving, and Holweger (1966), with subsequent modifications by M. Lemke), along with model
atmospheres computed by ATLAS9 (Kurucz 1997). Our spectra are sufficiently uncrowded that we can simply
compute abundances from equivalent widths, instead of performing a full spectral synthesis fit. Only lines attributed to a single
chemical species were considered; potentially blended lines are ignored in this analysis. Microturbulent velocity
$\xi$ was chosen such that the abundance derived for a single species (Fe II) was invariant with $\Weq$. We 
assumed a cluster metallicity of [Fe/H]~$= -1.5$~dex 
in computing the model atmospheres, and although many of the stars turn out to be considerably more metal-rich
than this (see below), adjustments to the atmospheric input were found to have only modest effects ($< 0.2$ dex) on 
the abundances of individual elements. Table 1 lists the final photospheric parameters used for
each of the target stars, as well as the heliocentric radial velocities.


\section{Results}

In Figure 1, abundance determinations for key chemical species are plotted as a function of stellar $\Teff$.
Note that the bottom three panels have a different vertical scale than the top six.
The values [X/H] represent logarithmic offsets from the solar values of Anders \& Grevesse (1993).
Whenever possible, we used the abundance computed for the dominant ionization stage of each element, to minimize the
possibility of non-LTE effects. The error bars incorporate the scatter among multiple lines of the same species, plus 
the uncertainties in $\Teff$, $\logg$, $\xi$, $\Weq$ for each line, and [Fe/H] of the input atmosphere. Even with the
conservative error bars in $\Teff$ ($\pm 500$ K in the cooler stars) and $\logg$ ($\pm 0.4$ dex), individual
element abundances are uncertain by 0.3 dex or less, with the sole exception of the Ca I lines of star IV-83.

The abundances of helium, iron, and magnesium provide the most striking contrast in behavior. The He
abundance first appears at the expected solar He/H ratio at $\Teff \simeq 11000 \unit{K}$, but then drops by
a factor of more than 100 as $\Teff$ increases to $19000 \unit{K}$. Iron, similarly, is present at (or slightly below) the
[Fe/H]~$=-1.51$~dex expected for this metal-poor cluster, but then rises to Population I abundances for the
stars hotter than $\simeq 12000$ K. Magnesium, on the other hand, appears consistently at almost exactly the canonical cluster
metallicity, with no discernable change with $\Teff$.

Other metals exhibit similar enhancements in the hotter stars. The Ti abundance rises by approximately a
factor of 30 from 8000 K to 15000 K, although the trend is not as clear-cut as with iron. Silicon and calcium are also
modestly enhanced to [X/H]~$\simeq -0.5$~dex among some of the hotter stars. The most pronounced overabundances are seen in
phosphorus, which appears at [P/H]~$\simeq +1.5$~dex in six stars, and chromium, which climbs past
solar metallicity to reach a remarkable [Cr/H]~$= 3.10$~dex, an enhancement of more than a factor of $10^4$ over the metallicity of M13,
albeit in only one star. These values are each based on several separate
spectral lines, in close agreement with each other, so we are confident that they are not due to random errors or line
misidentification.

The CNO elements, particularly nitrogen, also show enhancements, although most of these abundances are based
on only a single line per species, and are therefore suspect. N~II appears in four of the hot stars, at [N/H] ranging
from $+1.6$ to $+3.5$~dex. Nitrogen enhancement from dredge-up of fusion-processed material is expected in evolved stars,
but not to this extent, so if these values are accurate, some other mechanism must be at work. A single oxygen line
appears at a more reasonable [O/H]~$= +1.0$~dex, and carbon is solar or slightly subsolar in three of the stars.


\section{Discussion}

An underabundance of helium on the BHB has been observed in several previous instances (Baschek 1975; Heber 1987; 
Glaspey \etal 1989, among others), and in fact appears to be typical for stars of this type. Michaud, Vauclair, and
Vauclair (1983, henceforth MVV), building on the original suggestion by Greenstein, Truran, and Cameron (1967), explain
the underabundances as a result of gravitational settling of helium, which can take place if the outer atmosphere of the star is
sufficiently stable. Our current results are significant in that they demonstrate a distinct trend in [He/H] with
$\Teff$ and $\logg$ along the HB, including cooler, lower-gravity stars with roughly solar helium abundances, such
that the magnitude of helium diffusion can be traced over a range of conditions. Although MVV do predict greater helium
depletion in their hotter, higher-gravity BHB models, the actual abundance pattern is also likely to depend on
stellar rotation rate. As pointed out both in MVV and Glaspey {\it et al.}, diffusion can easily be stymied by turbulence, mass
loss, and meridional circulation produced by stellar rotation. Many of the BHB stars in M13 are comparatively fast
rotators (Peterson \etal 1995), reaching $\vsini \simeq 40 \kms$, while the ten stars in our study with helium abundances
all exhibit extremely narrow lines, suggesting $\vsini < 6 \kms$ in most cases. A more comprehensive assessment of the
stellar rotations, and their potential effects on diffusion processes, will be presented in another paper (Behr \etal 1999a).

The MVV calculations indicate that helium depletion should be accompanied by photospheric enhancement of metals, as the
same stable atmosphere which permits gravitational settling also permits levitation of species with large radiative
cross-sections. Overabundances of factors of $10^3-10^4$ from a star's initial composition could be supported by
radiation pressure, although as Glaspey \etal hasten to point out, this is a ``necessary but not sufficient condition''
for actual abundance anomalies to appear, given the possibility of turbulent mixing or radiation-driven escape of metals
from the star. MVV make some initial assessment of the magnitudes of these variations, but future models will have to
explain more fully why some elements (N, P, and Cr) are enhanced so much more strongly than others (Fe, C, Ti, Ca, Si),
while a few (Mg) are apparently immune to diffusion mechanisms. None of the recent diffusion work reported in the literature
treats the specific case of the BHB, so a detailed comparison of our results with current theory will have to wait for improved
models.

Our results for iron and magnesium closely parallel those of Glaspey {\it et al.}, who studied
two stars in globular cluster NGC 6752, one at 10000 K, the other at 16000 K. The hotter star displays a Fe enhancement
of 50 times above the cluster mean, but the cooler star has the same [Fe/H] as the cluster, while the Mg abundances are
near the cluster mean in both cases. They find significantly lower amounts of silicon and phosphorus than we do, but the rough
agreement in the helium and iron anomalies between these two BHB stars in NGC 6752 and our larger sample in M13
suggests that these diffusion mechanisms are not peculiar to M13.

In addition to furthering our understanding of diffusion mechanisms, these atmospheric abundance variations offer
potential ramifications for the photometric morphology of globular clusters. A great deal of recent attention has
focused on the presence of gaps in the color distribution of stars on the HBs of M13 and several other clusters (Ferraro
\etal 1997b; Sosin \etal 1997; Buonanno \etal 1986). The origin of such gaps is not yet understood, and presents
a challenge for theories of HB evolution. A prominent gap in M13's BHB, located at $\Teff \simeq 11000 \unit{K}$ 
and labelled `G1' in Ferraro \etal (1997b), seems to coincide with the onset of our diffusion anomalies. We will explore 
this possible connection in later publications. 

Additionally, there is the issue of ``overluminous'' regions of the BHB in several
clusters including M13 (Moehler 1998; Grundahl, VandenBerg, \& Anderson 1998; Grundahl \etal 1999). While cooler BHB stars are in good
agreement with theoretical zero-age horizontal-branch (ZAHB) tracks, those in the range $11000 \unit{K} \leq \Teff \leq
20000 \unit{K}$ are found to be significantly brighter (or equivalently, at lower $\logg$) than expected. Again, this is
the temperature range where diffusion effects start to alter the atmospheric composition significantly, possibly also
modifying the atmospheric structure. The abundances that we observe should prove useful in evaluating the potential role 
of diffusion-driven metal enhancement in explaining this phenomenon. 

Lastly, the observed helium diffusion may
have some impact on estimates of globular cluster ages. Helium diffusion in main-sequence models can alter evolutionary
timescales by 10\% or more (VandenBerg, Bolte, \& Stetson 1996), and although the atmospheric structure of HB stars and
MS stars are quite different, the magnitude of helium diffusion seen at different $\Teff$ and $\logg$ on the HB may offer some
insights into the degree of diffusion expected in the main-sequence case. More importantly, the helium fraction can
influence the luminosity of the ZAHB (Proffitt 1997), which will affect age determinations based on the $\Delta V$
between the turnoff and the HB, as well as distance estimates using the observed magnitudes of HB stars.

Further observational and theoretical work will be necessary to determine what relationship (if any) exists 
between the onset of these diffusion-driven abundance anomalies and other characteristics of the HB, such as its 
luminosity, stellar rotation, and gaps in its color distribution, and whether diffusion significantly affects estimates of GC ages.


\clearpage
\acknowledgments

These observations would not have been feasible without the HIRES spectrograph and the Keck I telescope. We are indebted
to Jerry Nelson, Gerry Smith, Steve Vogt, and many others for making such marvelous machines, to the W. M. Keck
Foundation for making it happen, and to a bevy of Keck observing assistants for making them work. Patrick C\^ot\'e
graciously provided assistance with many of the HIRES observations. Thanks also go to Manuela Zoccali, Elena Pancino, and
Giampaolo Piotto for their reduction of the HST photometry, and to Michael Lemke for introducing us to the LINFOR package and
installing it locally. SGD was supported, in part, by the Bressler Foundation. This research has made use of the SIMBAD
database, operated at CDS, Strasbourg, France.



\clearpage
 
\figcaption[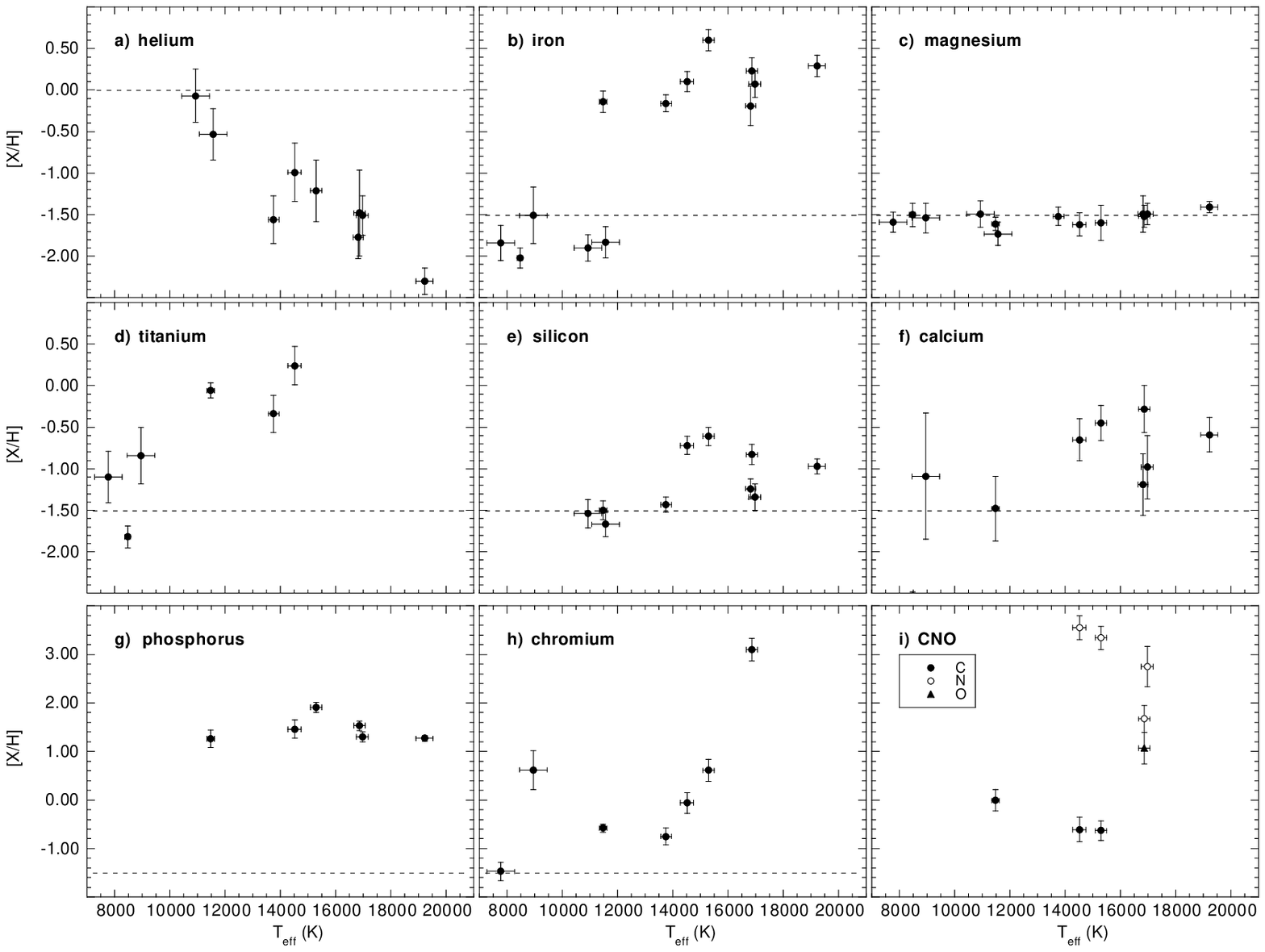]{He, Fe, Mg, Ti, Si, Ca, P, Cr, and CNO abundances over a range of $\Teff$, as log offsets from the
solar abundances. The dashed line represents the expected value of [X/H] for stars in M13 (ignoring
dredge-up of fusion-processed or $\alpha$-enhanced material) : $-1.51$~dex from solar for the metals, and the solar ratio for helium.
Note that the vertical scale is different in the bottom three panels.}


\clearpage 

\begin{deluxetable}{lrccc}
\tablenum{1}
\tablewidth{0pt}
\scriptsize
\tablecaption{Photospheric parameters for program stars}
\label{tab1}
\tablehead{
				&					&					&$\xi$ 				&Heliocentric	\nl
Star				&$\Teff$ (K)~~~		&$\logg$				&(km s${}^{-1}$)		&$v_r$ (km s${}^{-1}$)
}
\startdata
IV-83			&8960	\e 500		&3.37	\e 0.40		&3	\e 1				&$-253.4$	\nl
SA113			&11580	\e 500		&3.90	\e 0.40		&3	\e 1				&$-241.9$	\nl
SA404			&10940	\e 500		&3.78	\e 0.40		&1	\e 1				&$-250.8$	\nl
J11				&7780	\e 500		&3.10	\e 0.40		&3	\e 1				&$-246.7$	\nl
WF2--3035		&8470	\e 110		&3.26	\e 0.40		&2	\e 1				&$-253.7$	\nl
WF2--2541		&16970	\e 210		&4.72	\e 0.40		&1	\e 1				&$-257.9$	\nl
WF2--2692		&16860	\e 200		&4.71	\e 0.40		&2	\e 1				&$-236.2$	\nl
WF4--3085		&15300	\e 200		&4.50	\e 0.40		&0	$+\; 1$			&$-255.9$	\nl
WF2--820			&14520	\e 240		&4.39	\e 0.40		&0	$+\; 1$			&$-240.3$	\nl
WF3--548			&16820	\e 190		&4.70	\e 0.40		&2	\e 1				&$-235.7$	\nl
WF4--3485		&13760	\e 200		&4.27	\e 0.40		&0	$+\; 1$			&$-247.3$	\nl
WF3-1718		&11480	\e 140		&3.88	\e 0.40		&1	\e 1				&$-244.4$	\nl
WF2-3123		&19220	\e 310		&4.98	\e 0.40		&1	\e 1				&$-238.5$	\nl
\enddata
\end{deluxetable}


\end{document}